\documentclass[prl,twocolumn,superscriptaddress]{revtex4-1}
\usepackage{amsbsy}
\usepackage{amsfonts}
\usepackage{amssymb}
\usepackage{amsmath}
\usepackage{color}
\usepackage{graphicx}
\usepackage{float}
\usepackage{ulem}
\begin{document}

%opening

\title{Nonreciprocal Quantum Transport at Junctions of Structured Leads}

\author{Eduardo Mascarenhas}
\affiliation{Department of Physics and SUPA, University of Strathclyde, Glasgow G4 0NG, Scotland, United Kingdom}

\author{Fran\c{c}ois Damanet}
\affiliation{Department of Physics and SUPA, University of Strathclyde, Glasgow G4 0NG, Scotland, United Kingdom}

\author{Stuart Flannigan}
\affiliation{Department of Physics and SUPA, University of Strathclyde, Glasgow G4 0NG, Scotland, United Kingdom}

\author{Luca Tagliacozzo}
\affiliation{Department of Physics and SUPA, University of Strathclyde, Glasgow G4 0NG, Scotland, United Kingdom}
\affiliation{Departament de F\'{\i}sica Qu\`antica i Astrof\'{\i}sica and Institut de Ci\`encies del Cosmos (ICCUB), Universitat de Barcelona,  Mart\'{\i} i Franqu\`es 1, 08028 Barcelona, Catalonia, Spain}
\author{Andrew J. Daley}
\affiliation{Department of Physics and SUPA, University of Strathclyde, Glasgow G4 0NG, Scotland, United Kingdom}

\author{John Goold}
\affiliation{School of Physics, Trinity College Dublin, The University of Dublin, College Green
Dublin 2, Ireland}

\author{In\'es de Vega}
\affiliation{Physics Department and Arnold Sommerfeld Center for Theoretical Physics,
Ludwig-Maximilians-Universitat Munchen, D-80333 Munchen, Germany}

\begin{abstract}
We propose and analyze a mechanism for rectification of spin transport through a small junction between two spin baths or leads. For interacting baths we show that transport is conditioned on the spacial asymmetry of the quantum junction mediating the transport, and attribute this behavior to a gapped spectral structure of the lead-system-lead configuration. For non-interacting leads a minimal quantum model that allows for spin rectification requires an interface of only two interacting two-level systems. We obtain approximate results with a weak-coupling Born-master-equation in excellent agreement with matrix-product-state calculations that are extrapolated in time by mimicking absorbing boundary conditions. These results should be observable in controlled spin systems realized with cold atoms, trapped ions, or in electrons in quantum dot arrays. 

\end{abstract}

\maketitle

\textit{Introduction}. Recent experimental developments both in solid state and atomic physics have opened to opportunities to explore properties of quantum transport, both in and out of equilibrium ~\cite{Lattice,Cold,Cold2,Ions,Ions2,Dots,Dots2}. One important element of quantum transport is the possibility to generate rectification of currents, or non-reciprocal transport, that is, currents whose magnitude depend on the bias direction. Such phenomenon may arise from
\textit{asymmetry} and \textit{nonlinearity} in the underlying dynamics -- on a quantum mechanical level, effective nonlinearily is associated to interactions between particles.
In recent years, the study and design of systems that rectify transport has continuously expanded into the quantum regime, for example in optical systems~\cite{WaveDiode,EuValve,Filipo,EuInoOut,Rest1,Rest2,Rest3,Rest4,Rest5,Rest6,Rest7,Rest8,Rest9,Rest10,Rest11,Clerk,DiodeExp} and spin models~\cite{Zala,Landi1,Landi,Emmanuel,Emmanuel2,Emmanuel3,Emmanuel4,Valente,Nunnenkamp}. Most of the recent literature has focused on phenomenological Markovian baths. In this regard, spin rectification was demonstrated for XXZ models with asymmetric coupling along the Z axis \cite{Zala,Landi1,Emmanuel3}. The system can be mapped to spinless interacting fermions  and in this setting  the ZZ couplings correspond to denisty-density interactions. The presence of such interactions is believed to be paramount for the presence of rectification of the spin current. Although non-markovian features of quantum transport have been shown in several works~\cite{CountingSaro,Counting2,eWaiting,Counting3,China,China2}, rectification has not been analysed in this context.

\begin{figure}
{\includegraphics[width = 8cm]{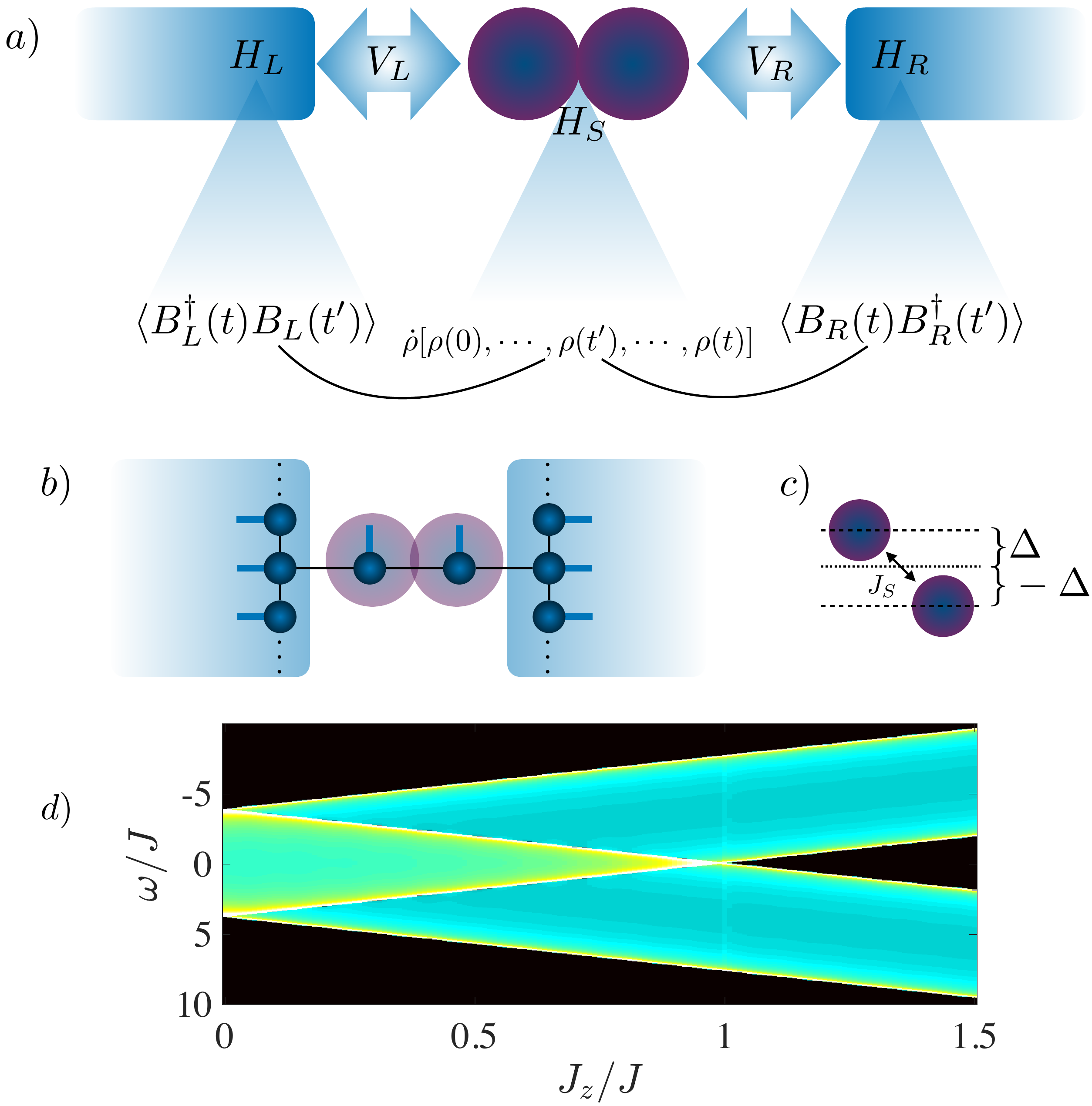}}\\ 
\caption{ $a)$ Representation of bath-system-bath coupling and effective evolution of the central interface that depends on its past history and time correlations of the bath.
b) The system-lead state $|\Psi\rangle$ depicted as a matrix-product-state reflecting the corresponding Hamiltonian structure of bulk-coupling. c) The small asymmetric interface between the leads is represented as a two qubit structure. d) The system-bath asymptotic nonequilibrium spectral function $A_{\Pi}(\omega)$ under the Born approximation for different $J_z$ couplings showing the spectral structure of the bath with a gap around $\omega=0$ for certain regimes in the limit $H_S\ll H_B$ ($J_S,\gamma=0.01J$).}
\label{Figure1}
\end{figure}

We adopt an open system approach beyond the Markov approximation to analyze a nonequilibrium many-body problem in which two interacting leads modeled as XXZ spin chains are coupled by a small interface, the open system, as depicted in Fig.~\ref{Figure1}-$a)$. For a non-interacting interface, we present a mechanism for rectification of spin currents, which arises from the \textit{spectral structure} of the leads (see Fig.~\ref{Figure1}-$d)$). We also show that for structureless leads a non-interacting system is fully reciprocal in agreement with the previous literature. Conversely, for structureless baths a minimal model allowing for efficient rectification requires a junction of two \textit{interacting} spins.

Interestingly, we show that the Born weak-coupling master equation is in excellent agreement with matrix product state (MPS) simulations. The Born equation is determined by the first order correlations of the leads, which decay as power-laws in XXZ spin chains~\cite{CorrXXZ,CorrXXZedge}. Since such slow decay is a rather generic feature of many-body systems~\cite{Giamarchi}, the effects reported here could be observed in a number of different architectures such as optical lattices~\cite{Lattice}, cold atoms~\cite{Cold,Cold2}, trapped ions~\cite{Ions,Ions2}, and superconducting leads coupled to quantum dots~\cite{SuperLeads,Dots,Dots2}.

\textit{Model and transport}. We first discuss the general transport properties. As depicted in Fig.~\ref{Figure1}-a), we consider left and right leads coupled to an interface with a coupling Hamiltonian given by
 \begin{equation}V=V_L+V_R,\quad V_i=2\gamma \left[B_iS_i^{\dagger}+B_i^{\dagger}S_i\right],\label{interaction}\end{equation}
where $B_{L(R)}$ are left (right) bath operators at the junction coupled with the corresponding system operators $S_{L(R)}$. The system-bath coupling strength $\gamma$ is assumed to be small compared to the system's and baths' frequency scales. The global dynamics are then governed by $\dot{\rho}_{SB}=-i[H_S + H_{B}+V,\rho_{SB}]$,
with $H_S$ the system Hamiltonian and $H_{B}=H_{L}+H_{R}$ the sum of left and right lead Hamiltonians.

In an interaction picture we define $\widetilde{V}(t)=e^{iH_0t}Ve^{-iH_0t}$, with $H_0=H_S + H_{B}$ dictating the dynamics of the combined system-bath density matrix $\widetilde{\rho}_{SB}$. A system operator $O$ (that commutes with $H_S$ for simplicity) follows a continuity Heisenberg-equation from which we define the current operator at the left ($L$) system-bath junction $\frac{dO}{dt}\big|_L=j^{(O)}_L=i[V_L,O]$. Its average value may be written in a second order iterated form
\begin{equation}I_{L}(t)=\mathrm{tr}\left\{ \rho_{SB}(t)j^{(O)}_L\right\}=-i\int_0^{t}dt'\left\langle \left[\tilde{j}_{L}^{(O)}(t),\widetilde{V}(t')\right]\right\rangle_{t'},\label{current}\end{equation}
in which we have eliminated the first order term for convenience (this is exact for the cases we address here). A similar expression follows for the current at the right junction. Note that~(\ref{current}) is exact and requires solving the system-bath many-body dynamics since $\left\langle \cdots \right\rangle_{t'}=\mathrm{tr}\{ \cdots \rho_{SB}(t') \}$. We will focus on 1/2-spin transport such that $O=Z=S^{\dagger}S-SS^{\dagger}$ leading to the spin current operator $j_{L}^{(Z)}=j_L=i[ V_L,Z_L]=-i4\gamma\left[ B_LS_L^{\dagger} -S_L B_L^{\dagger}\right]$. %and $j_R=i4\gamma\left[ B_RS_R^{\dagger} -S_R B_R^{\dagger}\right]$.
Eq.~(\ref{current}) can be re-written as $I_{L}(t)=-i8\gamma^2\int_0^{t}dt' \Pi_{L}(t,t')$, in terms of the nonequilibrium two-particle retarded-Green's function~\cite{Mahan}
\begin{multline}
\Pi_{L}(t,t')=-i\Theta(t-t')\Big\langle \widetilde{B}_L(t)\widetilde{B}_L^{\dagger}(t')  \widetilde{S}_L^{\dagger}(t)\widetilde{S}_L(t')\\
 - \widetilde{B}_L^{\dagger}(t)\widetilde{B}_L(t')  \widetilde{S}_L(t)\widetilde{S}_L^{\dagger}(t')   +\widetilde{B}_L(t)\widetilde{B}_R^{\dagger}(t')  \widetilde{S}_L^{\dagger}(t)\widetilde{S}_R(t') \\
-\widetilde{B}_L^{\dagger}(t)\widetilde{B}_R(t')  \widetilde{S}_L(t)\widetilde{S}_R^{\dagger}(t')  + \mathrm{h. c.}
\Big\rangle_{t'}.\label{Pi}
\end{multline}
The total average current flowing through the system is defined as $I(t)=[I_L(t)-I_R(t)]/2=-i8\gamma^2\int_0^{t}dt' \Pi(t,t')$, with the total Green's function $\Pi=[\Pi_L-\Pi_R]/2$. 
The asymptotic current can be expressed as 
\begin{eqnarray}I(\infty)&=&-i\frac{8\gamma^2}{\sqrt{2\pi}}\int_{-\infty}^{\infty}dt' \int_{-\infty}^{\infty}d\omega e^{i\omega t'}\Pi(\omega)\nonumber \\
&=& 8\gamma^2 A_{\Pi}(0),\label{currentinf}\end{eqnarray}
where we have defined the joint nonequilibrium spectral function as $A_{\Pi}(\omega)=-\mathrm{Im}\left\{ \Pi(\omega) \right\}$
with $\Pi(\omega)$ being the Fourier transform of $\Pi(\infty,t')$ (in practice the asymptotic behavior is captured by $\Pi(t'+\infty,\infty)$).
Therefore, $A_{\Pi}(0)$ plays a similar role to a generalized zero frequency conductivity. 

\begin{figure}
{\includegraphics[width = 8cm]{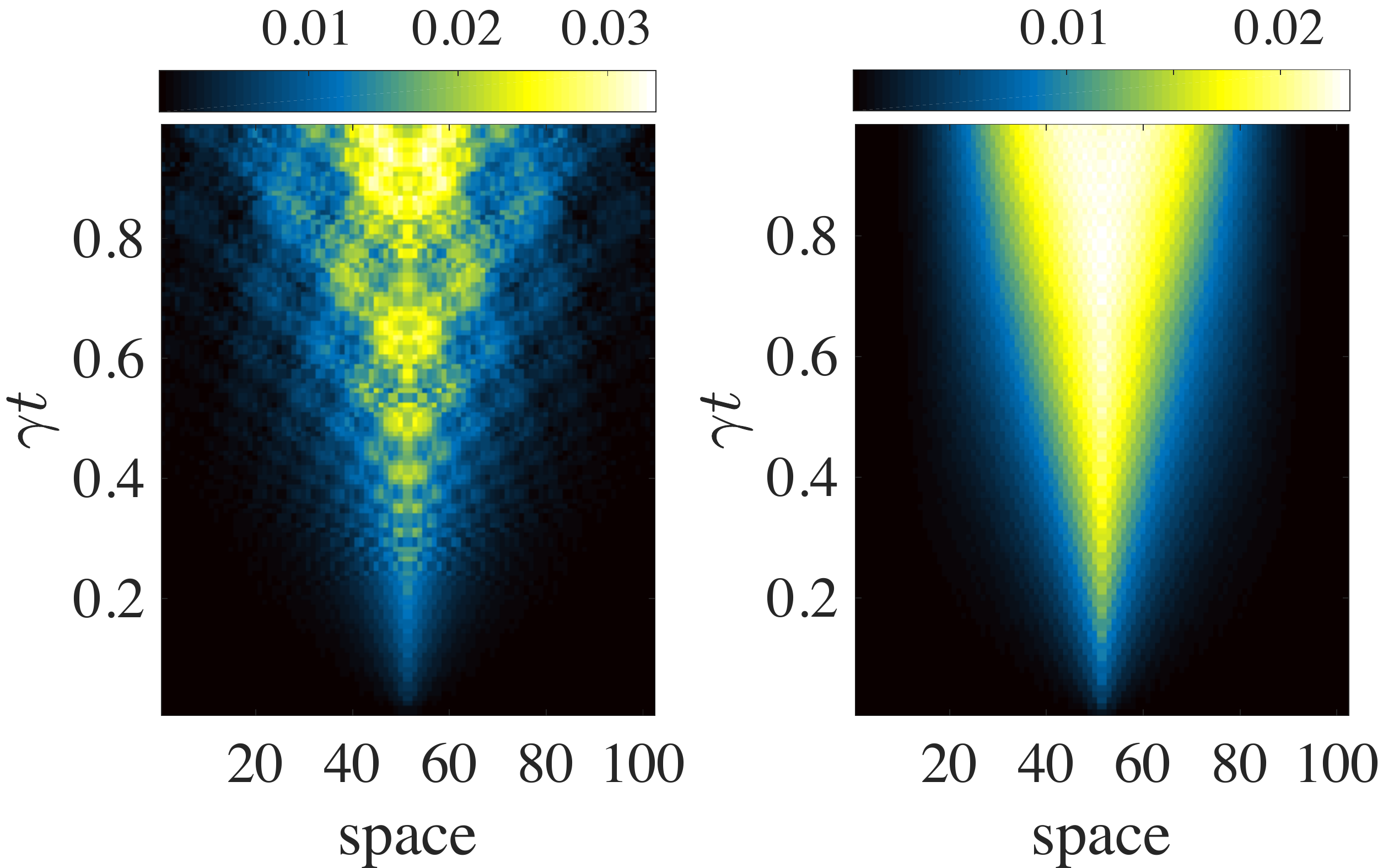}}\\ 
\caption{Currents at each bond in one of the leads without (left) and with (right) absorbing boundaries for $J_z=J$, $J_z^{\mathrm{system}}=0$ and $J_S=\Delta=\gamma=0.01J$. Reflection from the boundaries are suppressed in the right panel.}
\label{LightCones}
\end{figure}

\textit{Weak-Coupling approximations}. So far, the expressions derived for the current are exact. However, in order to proceed further it is convenient to consider approximations to the global system-bath state. One possibility is to use the Kubo approximation~\cite{Kubo1,Kubo2,Kubo3,Mahan} which assumes that the weak-perturbation only causes a small change in the global system-bath state $\widetilde{\rho}_{SB}(t)\approx \widetilde{\rho}_{SB}(0) +\gamma f(t)$. Although this is a good approach for the state of the baths, it might result as a rough approximation for the state of the small interface. 
The Born ansatz takes that into account and allows the system to evolve by considering that the global state is $\widetilde{\rho}_{SB}(t)\approx \widetilde{\rho}_{S}(t)\otimes \left[\widetilde{\rho}_{B}(0) +\gamma f(t)\right]$.
With Kubo's linear response theory~\cite{Kubo1,Kubo2,Kubo3,Mahan} 
the quantum average in~(\ref{Pi}) is substituted by $\left\langle \cdots \right\rangle_{00}=\mathrm{tr}\{ \cdots \rho_{SB}(0) \}$. 
Similarly, with the Born ansatz the average becomes $\left\langle \cdots \right\rangle_{t'0}=\mathrm{tr}\{ \cdots \widetilde{\rho}(t')\otimes\rho_{B}(0) \}$. These approximations assume that terms of order higher then $\gamma^2$ can be neglected. Also note that the Kubo approximation to the Green function~(\ref{Pi}) factors the first term (for example) into 
$\langle \widetilde{B}_L(t)\widetilde{B}_L^{\dagger}(t')\rangle_0 \langle \widetilde{S}_L^{\dagger}(t)\widetilde{S}_L(t')\rangle_0$, while the Born approximation leads to $\langle \widetilde{B}_L(t)\widetilde{B}_L^{\dagger}(t')\rangle_0 \langle \widetilde{S}_L^{\dagger}(t)\widetilde{S}_L(t')\rangle_{t'}$. Both approximations neglect correlations between system and bath, but also the correlations that emerge between different baths mediated by the system. Thus, the last two terms of~(\ref{Pi}) are dropped~\cite{Mark}.

\begin{figure}
{\includegraphics[width = 8cm]{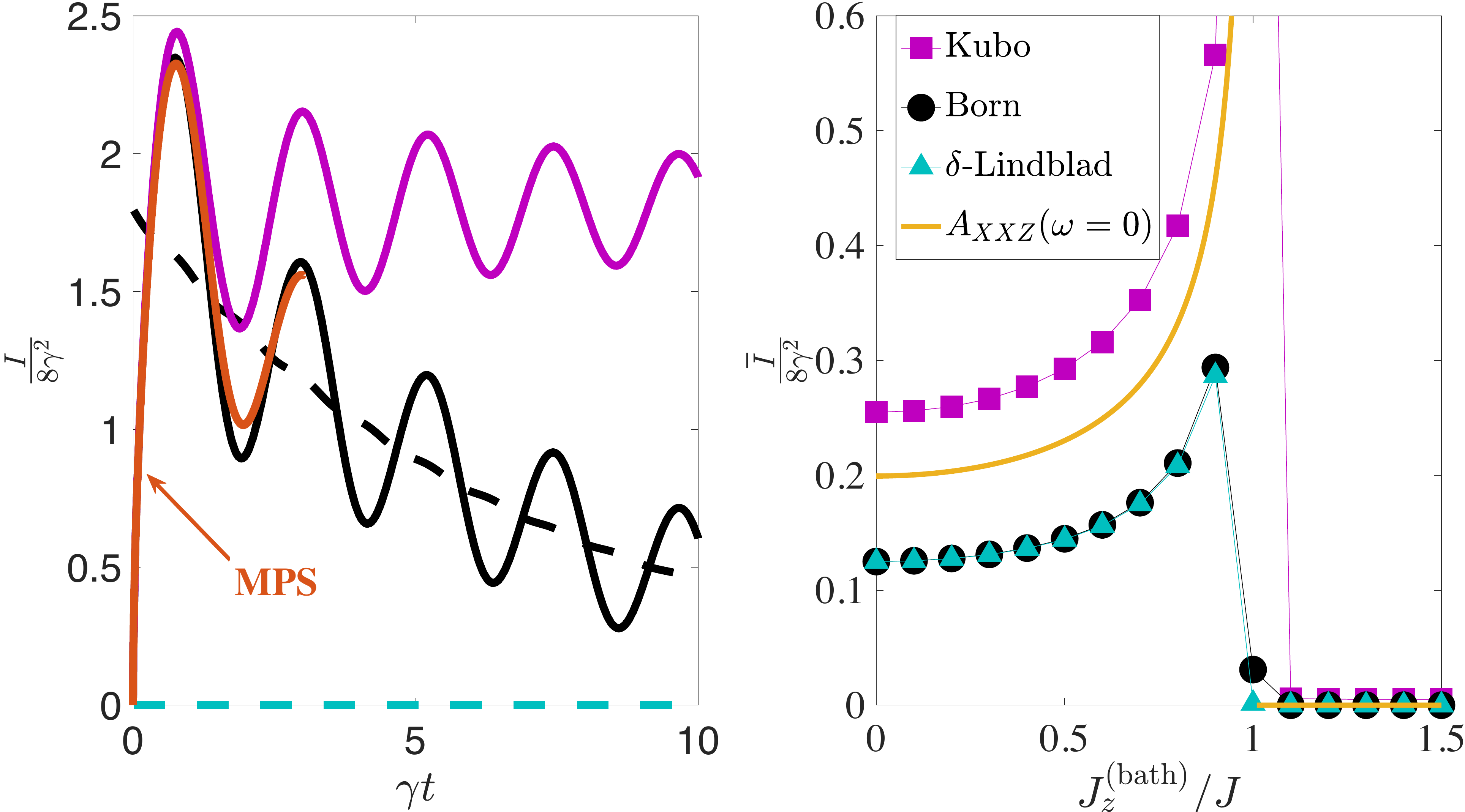}}\\ 
\caption{(Left) The MPS, Kubo and Born spin currents as a function of time at the Heisenberg point $J=J_z^{\mathrm{bath}}$. The dark-dashed line corresponds to~(\ref{GME}) and the light-dashed line to~(\ref{LME}). (Right) The corresponding asymptotic currents. Parameters are $J_z^{\mathrm{system}}=0$ and $J_S=\Delta=\gamma=0.01J$.}
\label{KuboBornLindblad}
\end{figure}

With the Born ansatz, tracing over the bath degrees of freedom we may compute the evolution for the system state~\cite{Carmichael,TOQS}. Dropping the subindex for the system state we have the Born-Master-Equation
\begin{multline} 
 \dot{\widetilde{\rho}}=-4\gamma^2 \int_0^t dt'\sum_{i} \Bigg[ \langle \widetilde{B}_i^{\dagger}(t)\widetilde{B}_i(t')\rangle \widetilde{S}_i(t)\widetilde{S}_i^{\dagger}(t')\widetilde{\rho}(t')  \\
 +\langle \widetilde{B}_i(t)\widetilde{B}_i^{\dagger}(t')\rangle \widetilde{S}_i^{\dagger}(t)\widetilde{S}_i(t')\widetilde{\rho}(t')
-\langle \widetilde{B}_i(t)\widetilde{B}_i^{\dagger}(t')\rangle \widetilde{S}_i(t')\widetilde{\rho}(t')\widetilde{S}_i^{\dagger}(t) \\
-\langle \widetilde{B}_i^{\dagger}(t)\widetilde{B}_i(t')\rangle \widetilde{S}_i^{\dagger}(t')\widetilde{\rho}(t')\widetilde{S}_i(t) + \mathrm{h. c.}
 \Bigg]
 \label{ME},
 \end{multline}
with the bath correlations dictating the dynamics of the system with memory on its past history as illustrated in Fig.~\ref{Figure1}-a).
Computing the long time evolution of the system state via~(\ref{ME}) can still be time consuming due to the integral-differential nature of the equation of motion and the power-law slowly decaying bath correlations. In the appendix we present a Redfield master equation directly targeting the Born steady state and an approximate approach that leads to a Lindblad form~\cite{Red,Lind1,Lind2,Lind3}.

\textit{Example of an XXZ bath}. We consider the leads (left and right) to be both described by XXZ spin-1/2 chains, with a Hamiltonian of the form
\begin{multline}\label{HE}
H_{XXZ}^{L(R)}=\sum_{r=-\infty}^{\infty} 2J\left[\sigma^{L(R)\dagger}_r\sigma^{L(R)}_{r+1} +\sigma^{L(R)\dagger}_{r+1}\sigma^{L(R)}_r \right]\\ +J_z^{(\mathrm{bath})}Z^{L(R)}_r Z^{L(R)}_{r+1},
\end{multline}
with $Z_r=\sigma_r^{\dagger}\sigma_r-\sigma_r\sigma_r^{\dagger}$, with $\sigma_r=|0\rangle_r\langle 1|_r$ and $J$ and $J_z$ the coupling strengths. 

The interface system is composed by two coupled asymmetric spins as represented in Fig.~\ref{Figure1}-$c)$, with the Hamiltonian
\begin{equation}H_S=2J_S(\sigma_L\sigma_R^{\dagger}+\sigma_L^{\dagger}\sigma_R)+\Delta(Z_L-Z_R)+J_z^{(\mathrm{system})}Z_LZ_{R},\end{equation}
where $\Delta$ is the relative detuning between the spins. As shown in Fig.~\ref{Figure1}-b), the left (right) system spin couples to a single spin of the left (right) lead. This is described with a Hamiltonian of the form (\ref{interaction}) with $S_{L(R)}=\sigma_{L(R)}$ and $B_{L(R)}=\sigma^{L(R)}_{0}$.
 
We analyze a \textit{nonequilibrium} protocol in which the left ($-$) and right ($+$) leads are prepared at zero temperature with a large bias $\pm \sum_r\mu Z_r$ added to their respective Hamiltonians while the system is initially prepared in an arbitrary state. Thus a global product state between the system and the leads is prepared with the leads oppositely maximally polarized. Then the bias is turned off and the global system is allowed to evolve generating spin currents.
In this limit, analytic results can be drawn under the first-order Holstein-Primakoff transformation for the baths~\cite{HP} $Z^{L(R)}_r=2[1-a^{L(R)\dagger}_ra^{L(R)}_r]$ and $\sigma^{L(R)}_r=-\sqrt{2}a^{L(R)\dagger}_r$. Under this approximation, the Hamiltonian of the baths take the quadratic form $H^{L(R)}_{XXZ}\approx -2\sum_{r}J[a^{L(R)\dagger}_ra^{L(R)}_{r+1}+\mathrm{h.c.}]+2J_za^{L(R)\dagger}_ra^{L(R)}_r$. The corresponding correlation function obeys $\langle B_L^{\dagger}(t)B_L(0)\rangle=e^{i4 J_z^{\mathrm{bath}}t}\mathcal{J}_0(4Jt)$,
with $\mathcal{J}_0$ being the zeroth-order Bessel function. 
This result was confirmed by exact MPS simulations (see appendix). 

Consequently, the decay rate that governs the relaxation dynamics of the interface can be written as $\Gamma(\omega)=\gamma^2\int_0^{\infty} d\tau e^{i\omega\tau} \langle B^{\dagger}_L(\tau)B_L(0)\rangle
=i\gamma^2[(4 J_z+\omega+i0^{+} )^2-(4 J)^2]^{-1/2}$, for each system transition of energy $\omega$. 
The decay rate may diverge, for example at $\omega=0$ and $J_z=J$ which could invalidate the perturbative approximation~\cite{Tudela}. The divergence can be avoided by slightly detuning the system away from the singularity. For the particular case addressed here, all transitions with $\omega=E'-E=0$ ($H_S|E\rangle=E|E\rangle$) are forbidden due to the symmetry $\langle E'|S_i|E\rangle=0$, thus ensuring the validity of the perturbative approach even at singular points.

\textit{Absorbing Boundaries}. Using matrix-product-state simulations represented in Fig.~\ref{Figure1}-$b)$, similar to~\cite{RecentQIM,MPS1,Luca1,Luca2,Schachenmayer,Anton,TDVP,TDVPprojector,VUMPS,IBMPS,Frauke1}, we find that the resulting dynamics are in great agreement with master equation for weak coupling. However, such regime produces very slow system relaxation that demands very large leads to prevent boundary reflections, rendering the simulations quickly unfeasible. To prevent this, we incorporate dissipation mechanisms in the baths mimicking absorbing boundary conditions (see appendix). This allows for a considerable extension of the time scales reachable by MPS simulations. The current-light-cones with and without the absorbing boundaries are compared in Fig.~\ref{LightCones}.

\begin{figure}
{\includegraphics[width = 8cm]{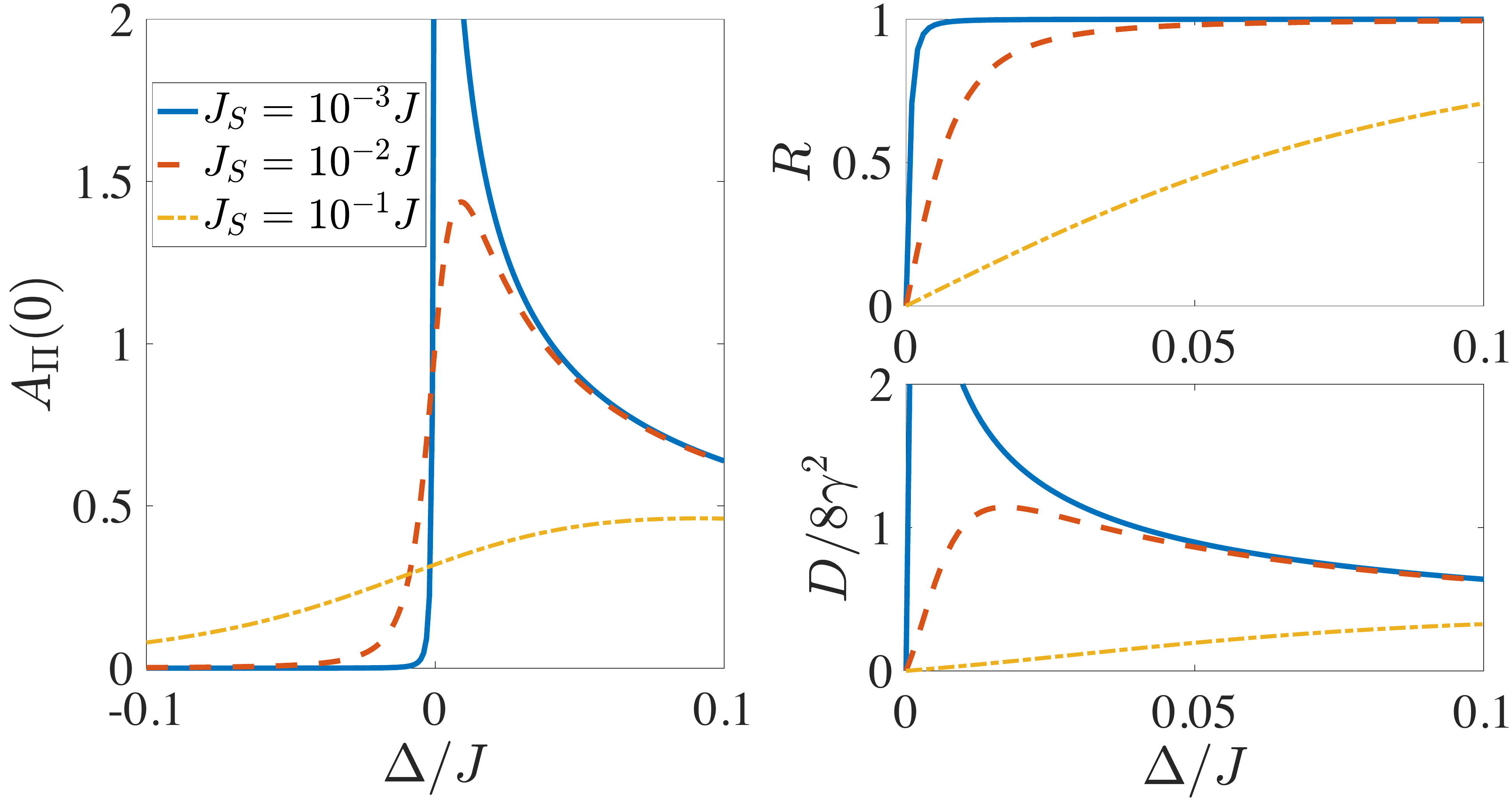}}\\
\caption{Analytical Kubo results for the short time response. (Left) The joint spectral function. (Upper-right) The rectification factor and (lower-right) the diode efficiency. Parameters are $J_z=J$.}
\label{KuboA&RBulkHighMuHpoint}
\end{figure}

\textit{Currents, Spectral Functions and Rectification}. When the system Hamiltonian is a small perturbation (the limit $H_S\rightarrow0$) the transport is largely governed by the physics of the bath. Generically, we expect ballistic transport for $J_z<J$ with non zero asymptotic currents, while diffusive or insulating behaviour is expected for $J_z>J$ with vanishing asymptotic currents. If $H_S=0$, the zero frequency spectral function follows the bath spectral function and we have $A_{\Pi}\approx A_{XXZ}$ with
$A_{XXZ}(\omega=0)=\mathrm{Re}\{[2\pi(J^2-J_z^2)]^{-1/2}\}/2$, for $ \mu\gg J$ which is shown in Fig.~\ref{KuboBornLindblad}. Thus $I(\infty)$ increases with $J_z^{(\mathrm{bath})}$ up to the Heisenberg point and then suddenly vanishes for $J_z^{(\mathrm{bath})}>J$ with the gap opening of $A_{XXZ}(\omega)$ around $\omega=0$. In Fig.~\ref{Figure1}-d) we show the full \textit{system-bath} asymptotic nonequilibrium spectral function $A_{\Pi}(\omega)$ under the Born approximation with the gap opening for $J_z^{(\mathrm{bath})}>J$ in agreement with the above analysis for the bath spectral function.

In Fig.~\ref{KuboBornLindblad} we analyze different approaches to describe the current for a weak system Hamiltonian. The MPS simulation shows an initial current burst that is also captured by the Kubo and Born approaches but not by the long time master equations. The relaxation of the system state after this burst is captured by the Born approach while ignored by Kubo. Although reliable, the Born evolution is time consuming. However, the steady state properties are captured by~(\ref{GME}) which amounts to a single algebraic equation to be solved. The long time currents are shown in Fig.~\ref{KuboBornLindblad} in agreement with the qualitative analysis of $A_{XXZ}$. The discrepancy between Kubo and the Born steady state results is significant whenever the currents are finite and is more drastic at the Heisenberg point, in which the spectral function is far from constant. 

The asymmetry parameter $\Delta$ of the system interface can induce non-reciprocal currents. To analyze this, we define the rectification associated to the total spin transported
\begin{equation} R_{\Delta}(T)=\frac{\overline{I}_{\Delta}(T)-\overline{I}_{-\Delta}(T)}{\overline{I}_{\Delta}(T)+\overline{I}_{-\Delta}(T)},\label{rect}\end{equation}
such that $\overline{I}_{\Delta}=\frac{1}{T}\int_0^TI_{\Delta}(t')dt'$ corresponds to the average current at a given asymmetry $\Delta$. There is usually a trade-off between the rectification factor and the current, in the sense that increasing the asymmetry might lead to higher $R$, however it usually also decreases the current since the sites become more and more out of tune~\cite{EuValve,Filipo,EuInoOut}. Thus as a last definition we have the diode factor $D_{\Delta}=\overline{I}_{\mathrm{sign}(R_{\Delta})\Delta}R_{\Delta}$ that captures this trade off and provides an overall fraction of the current that is rectified.

\begin{figure}
{\includegraphics[width = 8cm]{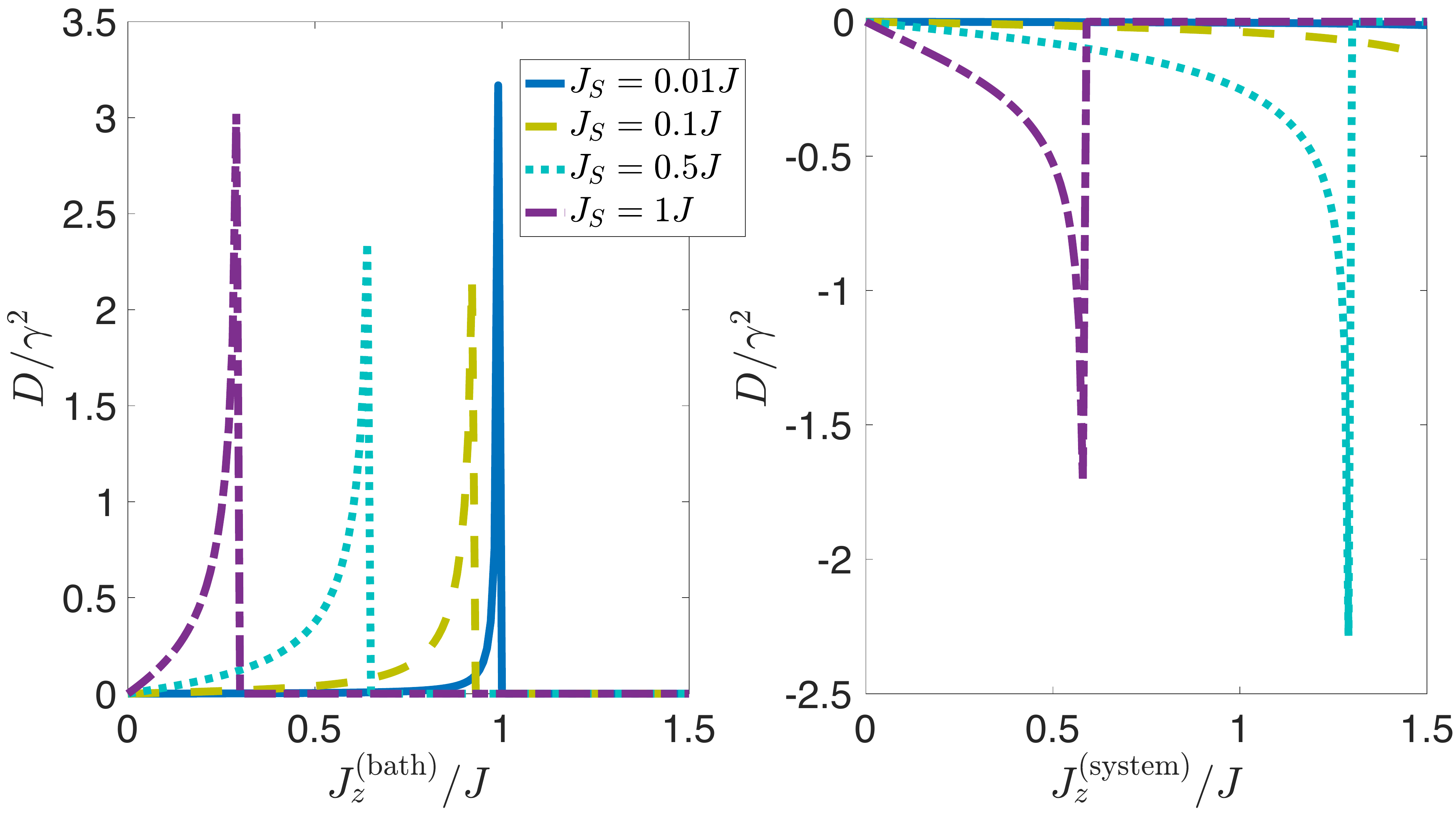}}\\ 
\caption{Asymptotic Born results for the diode factor for (left) a non-interacting system with interacting leads and (right) an interacting-system with non-interacting leads. We set $\Delta=J_S$.}
\label{SSR}
\end{figure}

As seen in Fig.~\ref{KuboBornLindblad} the simple Kubo approach provides an effective upper bound for the current, allowing for a qualitative description of the rectification mechanism. Analyzing the Kubo version of $A_{\Pi}$ at zero frequency and at the Heisenberg point we have a simple expression for the rectification factor
$R_{\Delta}^{\mathrm{Kubo}}(t\approx 1/\gamma)=\Delta[J_S^2+\Delta^2]^{-1/2}, \ \mathrm{for}\quad J_S,\Delta\ll J=J_z\ll\mu$.
In Fig.~\ref{KuboA&RBulkHighMuHpoint} we show that $A_{\Pi}(0)$ is asymmetric with respect to $\Delta$. In the perturbative limit $J_S\rightarrow0$, a positive $\Delta$ leads to a gapless spectral function, while a negative $\Delta$ leads to the open gap yielding perfect rectification $R=1$. A finite $J_S$ leads to a smooth crossover in $A_{\Pi}$ while still presenting finite rectification. In Fig~\ref{KuboA&RBulkHighMuHpoint} we also show the high rectification factor and the diode factor accounting for the trade off between asymmetry and total output current. 

The Kubo results are very accurate for short times, however, for long times we have to resort to the Born approach in Fig.~\ref{SSR}. For weak system Hamiltonian a rectification peak manifests just before the Heisenberg point (before the gap opening) as expected by the Kubo analysis (left panel of Fig.~\ref{SSR}). Increasing the magnitude of the system Hamiltonian shifts this peak towards lower bath-interactions indicating that the global spectral gap is shifted by the system Hamiltonian. Higher currents flow from the spin with positive frequency to the spin of negative frequency. In the opposite case of interacting interface and non-interacting leads the results are markedly different (right panel in Fig.~\ref{SSR}). The optimal transport direction is  inverted and higher currents flow from the negative-frequency spin to the positive one. 
We emphasize that the above results are only captured by the Born approach, which accounts for the dependency of the system dynamics on the spectral properties. In contrast, the local Lindblad equation only considers a single decay rate (i.e. decay channel) for the system and therefore fails for structured baths, particularly when the system frequencies are spread, i.e. when $J_S\approx J$ (see Appendix).

\textit{Conclusion}. In summary, we have analyzed in detail weak-coupling approximations for transport scenarios which are far from equilibrium showing how the Born approach goes well beyond linear-response and is in good agreement with exact MPS results. 
Considering a setting with a small system between two XXZ leads we have shown the presence of nonreciprocal transport. We have presented a mechanism for optimal rectification associated to the asymmetric spectral structure of system+bath induced by the system spacial imbalance. Our results indicate that phenomenological Lindblad approaches may fail since they do not take into account the spectral structure. Lastly, the mapping of the bath to a non-interacting model suggest that rectification may emerge for structured baths even in complete absence of interactions.

Acknowledgments: E.~M.~thanks Thierry Giamarchi, Ignacio Cirac, Mari-Carmen Banuls, Daniel Valente and Thiago Werlang for inspiring discussions. Work at the University of Strathclyde was supported by the EPSRC Programme Grant DesOEQ (EP/P009565/1), and by the EOARD via AFOSR grant number FA2386-14-1-5003. This work was supported by an SFI-Royal Society University Research Fellowship (J.G.). This project received funding from the European Research Council (ERC) under the European Union's Horizon 2020 research and innovation program (grant agreement No.~758403). I. D. V. was financially supported by the Nanosystems Initiative Munich (NIM) under project No. 862050-2 and the DFG-grant GZ: VE 993/1-1.  

\section{Appendix: Absorbing-Boundaries MPS simulations} 

\begin{figure}
{\includegraphics[width = 9cm]{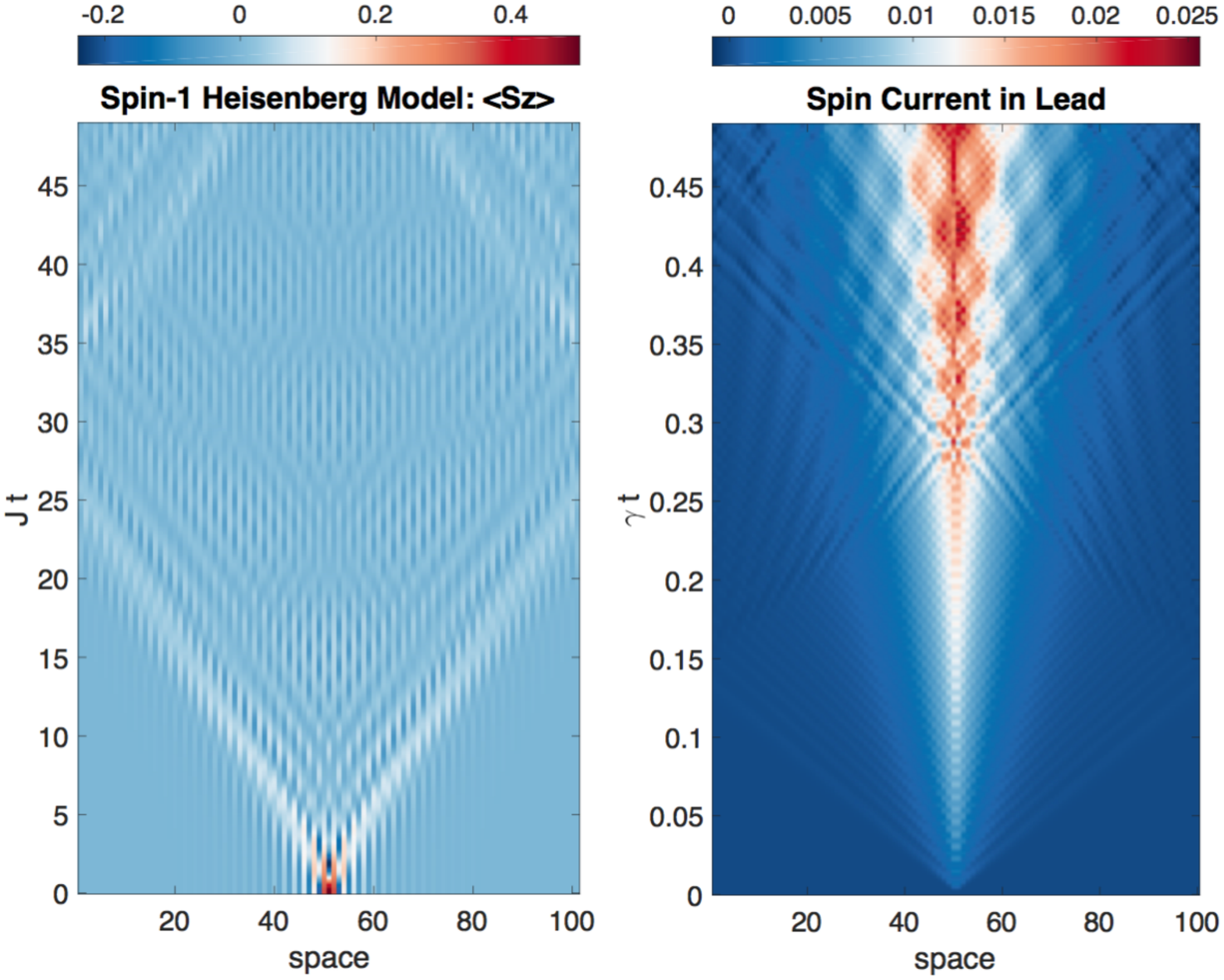}}\\ 
\caption{Infinite boundary MPS (IBMPS) results. Spin projection along z-axis for the antiferromagnetic spin-1 Heisenberg model of $100$ spins (left) after applying a spin creation operator at site 51, ($\sigma^{\dagger}_{51}$). Spin current at each bond in one of the leads for our open-system model (right). Reflection from the boundaries are delayed but not suppressed.}
\label{Inf_Bound}
\end{figure}

Simulating absorbing boundary conditions in classical physics is a relatively simple task typically accomplished by setting derivatives to zero at the boundary. In quantum mechanics this is still an open problem, in general, with some remarkable strategies for infinite-boundary-MPS simulations~\cite{IBMPS}. 
This strategy uses an infinite-MPS algorithm to compute the ground state, such as that presented in Ref.~\cite{VUMPS}. The authors of Ref.~\cite{IBMPS} then create a localized perturbation in the centre of the i-MPS and evolve the state using a finite-MPS time evolution algorithm. The only difference in the time evolution is how the sites at the boundary are treated. These need to be evolved by an effective Hamiltonian that takes into account the half infinite boundaries. 
In~\cite{IBMPS} the scheme is applied to the spin-1 Heisenberg model in the anti-ferromagnetic phase and they show that the perturbation is absorbed by the boundary, allowing them to evolve a smaller system for longer without worrying about boundary effects.   
However, this strategy fails for the setting considered here. In Fig.~\ref{Inf_Bound} (left) we apply this algorithm to the same system as the authors of Ref.~\cite{IBMPS}, i.e.\ the spin-1 Heisenberg model. The only difference is that we evolve the system with TDVP instead of TEBD. We show that while the perturbation is initially absorbed by the boundary, after evolving to longer times we do still see a reflected component. This delay in the reflection is not enough to allow us to evolve an open-system model for a significantly longer time. 

The situation gets worse when we apply this scheme to our open-system model, i.e two-sites coupled to two large leads where the leads are initially in a product state and the Hamiltonian is the spin-1/2 XXZ model. In Fig.~\ref{Inf_Bound} (right) we see that at the boundary of one of the leads, there is almost no delay in the reflection of the current, meaning that there is no advantage over simply using a finite-MPS algorithm.
The analysis indicates that the success of the strategy of Ref.~\cite{IBMPS} is highly dependent on the system and specific application. And while it indeed has great potential in some circumstances, as shown by their calculation of the lowest excitation branch of the spin-1 Heisenberg model~\cite{IBMPS}, it is not applicable in the context presented here. 
This analysis justifies the need to develop a different approach to absorbing boundary conditions.

We include dissipative processes into the original dynamics
\begin{equation}d|\Psi\rangle=\left[-iHdt-\sum_r \left(\frac{1}{2}J_r^{\dagger}J_rdt-J_rdQ_r\right) \right]|\Psi\rangle,\end{equation} 
with $H=H_S+H_B+V$ and $dQ_r=\langle J^{\dagger}_r+J_r\rangle dt+dW_r$ with $dW_r$ being Wiener process~\cite{Wiseman}. These dissipative processes are an auxiliary tool to absorb excitation in the bath and prevent their reflection at the boundary. The $J_r$ operators have to be chosen for the specific problem and there is no general recipe for constructing them. On the left lead, which is prepared in an all up state, we set $J_r=\sqrt{\zeta(r)}\sigma^{\dagger}_r$ and in the opposite side we set $J_r=\sqrt{\zeta(r)}\sigma_r$ with $\zeta(r)=e^{-\gamma_{B}r}$ with $r$ being the distance from the boundary and $\gamma_B$ the effective range of these dissipative processes.

\section{Appendix: Steady State master equations}
Assuming that the bath correlations do decay, no matter how slow, and that consequently the system reaches a steady state, we have the long time limit of the Born-master-equation \begin{multline} 
\dot{\rho}(t\rightarrow\infty)=0=-i\left[H_S,\rho\right]
-\sum_{s,\omega,\omega',\alpha,\alpha'} \Bigg[ \Gamma_{\alpha,\alpha'}^{(s)}(\omega') \\\times\left( K_{\alpha}^{(s)\dagger}(\omega)K_{\alpha'}^{(s)}(\omega')\rho- K_{\alpha'}^{(s)}(\omega')\rho K_{\alpha}^{(s)\dagger}(\omega)\right) + \mathrm{h. c.}
 \Bigg] \label{GME},\end{multline}
 with $\Gamma^{(s)}_{\alpha,\alpha'}(\omega)=\gamma^2\int_0^{\infty} d\tau e^{i\omega\tau} \left\langle \widetilde{X}^{(s)}_{\alpha}(\tau)\widetilde{X}^{(s)}_{\alpha'}(0)\right\rangle$, 
whose real part is an effective relaxation rate and imaginary part a system  frequency $\omega$~\cite{TOQS}.
We have assumed the eigen decomposition $H_S=\sum_E E P(E)$, $P(E)=|E \rangle\langle E|$ with $K^{(s)}_\alpha(\omega)=\sum_{E'-E=\omega} P(E)K^{(s)}_{\alpha}P(E')$, $K^{(s)}_{\alpha}= i^{\alpha}\left[ S_s^{\dagger}+(-1)^{\alpha}S_s \right]$ and $X_{\alpha}^{(s)}=(-i)^{\alpha}\left[ B_s+(-1)^{\alpha}B_s^{\dagger} \right]$ with $\alpha=0,1$. The master equation~(\ref{GME}) is commonly referred to as the \textit{global} approach since it contains the $K$ operators which can be delocalized in space, however it is not of Lindblad form since we have not discarded terms that couple different frequencies $\omega\neq \omega'$. Alternatively, \textit{phenomenological} assumptions of the form $\langle \widetilde{B}_i^{\dagger}(t)\widetilde{B}_i(t')\rangle\approx \Gamma^{(i)}_h\delta(t-t')$ and $\langle \widetilde{B}_i(t)\widetilde{B}^{\dagger}_i(t')\rangle\approx \Gamma^{(i)}_p\delta(t-t')$ lead to the zero frequency \textit{local} approach 
\begin{multline} 
\dot{\rho}_{t\rightarrow\infty}\approx-i\left[H_S,\rho\right]
-\sum_{s} \Bigg[ \Gamma_{p}^{(s)}\left( S_s^{\dagger}S_s\rho- S_s\rho S_s^{\dagger}\right) + \mathrm{h. c.}
 \Bigg]\\
 -\sum_{s} \Bigg[ \Gamma_{h}^{(s)}\left( S_sS_s^{\dagger}\rho- S_s^{\dagger}\rho S_s\right) + \mathrm{h. c.}
 \Bigg]
  \label{LME},\end{multline}
 with $\Gamma^{(s)}_{h}=4\gamma^2\int_0^{\infty} d\tau \langle \widetilde{B}_s^{\dagger}(\tau)\widetilde{B}_s(0)\rangle$ and $\Gamma^{(s)}_{p}=4\gamma^2\int_0^{\infty} d\tau \langle \widetilde{B}_s(\tau)\widetilde{B}_s^{\dagger}(0)\rangle$. Note that $\Gamma^{(s)}_{h}$ and $\Gamma^{(s)}_{p}$ can be expressed in terms of the zero frequency $\Gamma^{(s)}_{\alpha,\alpha'}(\omega=0)$.
Hence, the phenomenological Lindblad approach described in (\ref{LME}) is only accurate when the system frequencies are not spread in comparison to the bath spectral function, in such a way that the decay rates corresponding to each system decay channel $\Gamma^{(s)}_{\alpha,\alpha'}(\omega)$ in Eq. (\ref{GME}) can be well approximated by a single one  $\Gamma^{(s)}_{\alpha,\alpha'}(\omega)\approx\Gamma^{(s)}_{\alpha,\alpha'}(\omega=0)$. Taking into account the multiple decay channels of the system appears to be crucial not only to describe transport properties, as described here, but also to describe thermalization~\cite{Red,Lind1,Lind2,Lind3}.

\section{Appendix: Time Correlations}

Let us start by determining the bath correlations beginning with cases that can be quickly solved analytically, that is, the non-interacting XX model with $J_z=0$~\cite{Giamarchi}. To describe the bulk physics we may assume periodic boundary conditions and perform the Jordan-Wigner transformation $\sigma_x=e^{i\pi \sum_{x'=0}^{x-1}c^{\dagger}_{x'}c_{x'}}c_x$, shift the momentum of the fermions by $\pi$ as $c_x\rightarrow(-1)^{x}c_x$ and perform a Fourier transformation $c_x=\frac{1}{\sqrt{N}}\sum_{k=0}^{N-1}q_ke^{i2\pi xk/N }$ leading to the decoupled representation of the Hamiltonian $H_{XX}=\sum_k \omega_kq^{\dagger}_kq_k$ with $\omega_k=-4J\cos(2\pi k/N)$ the usual tight-biding or free particles on a 1D lattice dispersion relation. 
The correlations are then given by 
\begin{eqnarray}G_{XX}^{\mathrm{bulk}}(t,\beta,\mu)&=&\langle \sigma^{\dagger}(t)\sigma(0)\rangle=\lim_{N\rightarrow\infty}\frac{1}{N}\sum_kn(\omega_{k})e^{i\omega_kt} \nonumber \\
&=& \frac{2}{N}\int_{-4J}^{4J}\frac{dk}{d\omega}n(\beta,\omega,\mu)e^{i\omega t}d\omega 
,\end{eqnarray}
with $\frac{dk}{d\omega}=\frac{N}{8J\pi}\left[ 1-\left(\frac{\omega}{4J}\right)^2\right]^{-1/2}$ and $n(\beta,\omega_k,\mu)=\mathrm{tr}\left\{ q_k^{\dagger}q_k\rho_E \right\} =\left[1+e^{\beta(\omega_k-\mu)}\right]^{-1}$ assuming an initial equilibrium state $\rho_E\propto e^{-\beta(H-\mu\sum_xZ_x/2)}$. At zero temperature the mode occupation tends to the Heaviside step function $n(\beta\rightarrow\infty,\omega,\mu)=\Theta\left[-(\omega-\mu) \right]$, thus we omit $\beta$ in the following. In this limit, we can easily determine the correlations and their asymptotic ($t\gg1$) forms collecting only the leading (slowest) power law. 

\begin{figure}
{\includegraphics[width = 8cm]{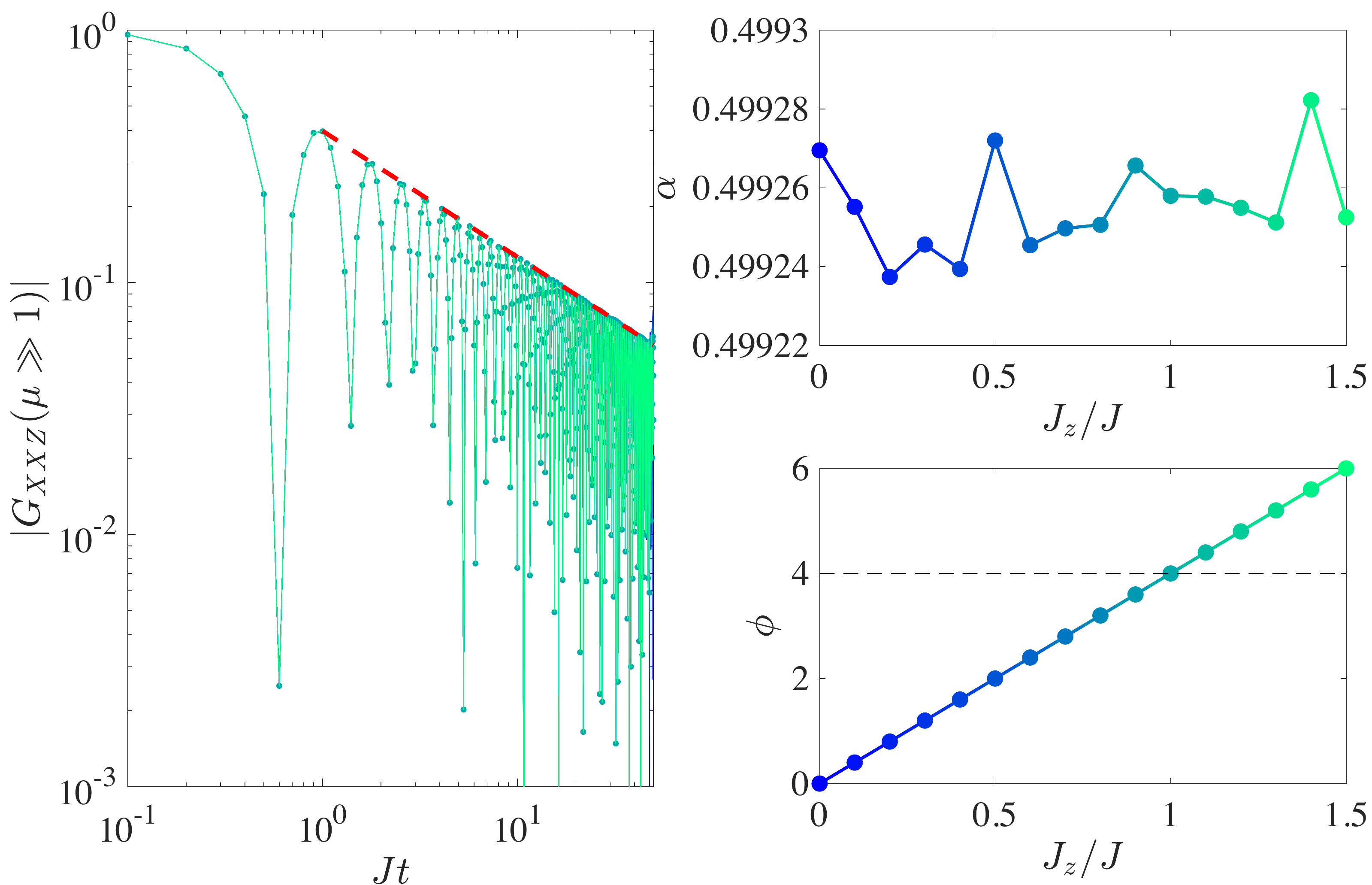}}\\ 
\caption{ MPS results for correlations with a chain of 200 spins in agreement with the analytical results. Darker colors correspond to lower $J_z$ while brighter colors correspond to higher $J_z$. The modulus of the correlation in log-log scale for (Upper left) $\mu\gg1$. The red dashed line is the asymptotic expression in~(\ref{1/2}), respectively. The Bessel functio result is indistinguishable from the MPS results. (Upper right) The fitted power law exponent and (lower right) the corresponding phase frequency assuming $G_{XXZ}\sim e^{i\phi t}t^{-\alpha}$. Thus we have confirmation that $\alpha=1/2$ and $\phi=4J_z$.}
\label{TcorrEdge}
\end{figure}

For large $\mu$ we have 
\begin{equation}G_{XX}^{\mathrm{bulk}}(t,\mu\gg1)=\mathcal{J}_0(4Jt) \sim \frac{ \cos[\pi/4 - 4 Jt ]}{ \sqrt{2 \pi Jt} } \label{1/2}
\end{equation}
such that $\mathcal{J}_n$ is the $n$-th order Bessel function. 

Reincorporating interactions into the model we opt to treat the large $\mu$ and arbitrary $J_z$ regime via a first order Holstein-Primakoff transformation~\cite{HP} $Z_x=2[1-a_x^{\dagger}a_x]$ and $\sigma_x=-\sqrt{2}a_x^{\dagger}$. Under this approximation, the Hamiltonian takes the form 
\begin{equation} H_{XXZ}\approx -2J\sum_{x}[a_x^{\dagger}a_{x+1}+\mathrm{h.c.}] -4J_z\sum_{x}a_x^{\dagger}a_x,\end{equation}
with a clear interpretation of $J_z$ as a local potential or detuning with the edge sites receiving only half of the detuning of the bulk sites since they have only one neighbour. Applying the free particle techniques outlined previously we find
\begin{equation}G_{XXZ}^{\mathrm{bulk}}(t,\mu\gg1)=e^{i4J_zt}G_{XX}^{\mathrm{bulk}}(t,\mu\gg1).\end{equation}

The system equilibrium correlations can also be obtained by similar techniques. We have systematically checked that our conclusions are not dependent on the initial state of the system. Therefore, for simplicity we choose to work with an initially down polarized $|\downarrow\downarrow\rangle$. The system correlation of interest is then 
\begin{multline}\langle \widetilde{S}_L(t)\widetilde{S}_L^{\dagger}(0)\rangle=\cos \left(2 t \sqrt{\Delta ^2+J_S^2}\right)\\
-\frac{i \Delta  \sin
   \left(2 t \sqrt{\Delta ^2+J_S^2}\right)}{\sqrt{\Delta ^2+J_S^2}}.\end{multline}

\end{document}